\documentclass[preprint,preprintnumbers,amsmath,amssymb]{revtex4}
\usepackage{graphicx}
\usepackage{epsfig}
\usepackage{lscape}
\usepackage{dcolumn}
\usepackage{bm}
\begin{document}

\title{ Time dependent density functional theory calculation of
van der Waals coefficient C$_{6}$ of alkali-metal atoms Li, Na, K, alkali
dimers Li$_{2}$, Na$_{2}$, K$_{2}$ and sodium clusters Na$_{n}$ }

\author{Arup Banerjee$^{a}$ and  Jochen Autschbach$^{b}$,\\
(a) Laser Physics Application Division, Raja Ramanna Centre for Advanced Technology\\
Indore 452013, India\\
(b) Department of Chemistry, State University of New York at
Buffalo, Buffalo, New York 14260-3000}

\begin{abstract}
In this paper we employ all-electron time dependent density
functional theory (TDDFT) to calculate the long range
dipole-dipole dispersion coefficient (van der Waals coefficient)
$C_{6}$ of alkali-metal atoms Li, Na, K, alkali-metal atom dimers
Li$_{2}$, Na$_{2}$, K$_{2}$ and sodium clusters containing even
number of atoms ranging from 2 to 20 atoms. The dispersion
coefficients are obtained via Casimir-Polder expression which
relates it to the frequency dependent linear polarizabilty at
imaginary frequencies.   The frequency dependent polarizabilities
are calculated by employing TDDFT--based complete sum-over-states
expressions for the atoms, and direct TDDFT linear response theory
for the closed shell dimers and clusters.
\end{abstract}

\maketitle

\section{Introduction}
The contribution of the long-range van der Waals force (dispersion
force) to the interaction between two many-electron systems is
quite significant. This force plays an important role in the
description many physical and chemical phenomena such as adhesion,
surface tension, physical adsorption etc. Physically, this long
range force arises from the correlation between the electron
density fluctuations at widely separated spatial locations such
that the electrons belonging to the different molecules are
distinguishable. For such a large separation a mathematical
expression for the potential corresponding to the long range
dispersion force is obtained by employing perturbation theory for
the calculation of the second-order change in energy due to
coulomb interaction between two charge distributions. The first
term in the perturbative expansion of the interaction potential
(after orientational averages have been performed) decays as
$-C_{6}/R^{6}$, where R is the intermolecular distance and the van
der Waals coefficient $C_{6}$ describes the dipole-dipole
interaction between the two polarizable systems. The calculation
of this coefficient can be performed by using the Casimir-Polder
expression \cite{casimirpolder,stone} which relates it to the
dynamic dipole polarizability at imaginary frequencies. The
dynamic polarizability which describes the response of an atom or
molecule to a weak, time dependent external electric field has
been well studied. There exist a number of wavefunction--based
ab-initio methods for calculating this quantity at varying levels
of sophistication taking into account electron correlation.
Alternatively, time dependent density functional theory (TDDFT)
represents an efficient tool for first--principles theoretical
calculations of dynamic polarizability of atoms and molecules,
typically at significantly lower computational cost as correlated
wavefunction--based methods. TDDFT often yields an accuracy
similar to wave function based correlated methods as long as
long--range charge transfer excitations are not overly important
for the property under consideration. Recently, TDDFT has been
applied to calculate van der Waals coefficients of variety of
atoms and small molecules \cite{gisbergen1} extensive set of
polycyclic aromatic hydrocarbons \cite{jiemchooroj1,miugel},
C$_{60}$ and C$_{70}$ molecules \cite{jiemchooroj1,banerjee1} and
small sized sodium clusters up to 20 atoms
\cite{jiemchooroj2,banerjee2}. The results of these calculations
are quite encouraging and clearly demonstrate that TDDFT based
method yields results for C$_{6}$ which are very close to other
theoretical and experimental data (where available). This success
has motivated us to further apply all-electron TDDFT based method
to calculate the coefficient C$_{6}$ to variety of systems
containing alkali-metal atoms.

In this paper, we calculate C$_{6}$ for following interactions (i)
alkali atom-alkali atom (Li, Na, and K), (ii) alkali atom-alkali
atom dimer (Li$_{2}$, Na$_{2}$, and K$_{2}$), and (iii) alkali
atom-sodium cluster (Na$_{n}$,where $n\leq 20$). The choice of
alkali-metal atoms and their molecules for our calculations are
also motivated by the recent development of laser cooling and
trapping of these atoms \cite{atomcooling}. These advances have
rekindled the interest in the knowledge of long range forces
between alkali atoms and clusters thereof as these forces play an
important role in the properties of cold gases of atoms and
molecules. Moreover, the results of the calculations of C$_{6}$
for systems (i) and (ii) give a good opportunity to assess the
accuracy of TDDFT to calculate C$_6$ for alkali atoms and diaomics
as for these systems accurate theoretical data obtained by
correlated wave-function based methods exist in the literature.
Moreover, the calculations of C$_{6}$ for the alkali atom --
Na$_{n}$ interactions is motivated by an experiment performed by
Kresin and co-workers \cite{kresin1}. Kresin et al.\ measured the
integral scattering cross section in low energy collisions between
a beam of sodium clusters and vapours of alkali-metal atoms Li,
Na, and K. For low energy collisions, the integral scattering
cross section depends on the van der Waals coefficient C$_{6}$.
The experimental results for integral scattering cross section
matched quite well with the theoretical predictions which were
obtained by employing the London dispersion formula for $C_{6}$.
The London dispersion formula assumes that all the strength of dipole
transition is concentrated in a single peak located at an effective frequency
and the formula involves only two parameters. (see Eq. (\ref{london}) below). We will compare our TDDFT based results with the numbers used in Ref. \cite{kresin1}
to reproduce their experimental data. We note that, to the best of
our knowledge, no first--principles theoretical data for C$_{6}$
between alkali-metal atom and sodium clusters have so far been
reported in the literature.

Before we proceed with the main plan of the paper, we wish to
mention here that in principle ground state density functional
theory (DFT) should yield the exact ground-state properties
including the long range van der Waals energies. However, the
widely used local density approximation (LDA) and generalized
gradient approximations (GGA) \cite{gga1,gga2,gga3}
exchange-correlation (XC) functionals as well as popular hybrid
functionals fail to reproduce the van der Waals energies. This is
due to the fact that the LDA and GGA functionals cannot describe
the correlated motion of electrons arising from Coulomb
interaction between distant non overlapping electronic systems. It
is only recently that attempts
\cite{andersson,dobson,kohn,becke,olasz} have been made to obtain
van der Waals energies directly from the ground-state energy
functional through systematic improvements of the effective
Kohn-Sham potential. On the other hand, it is possible to make
reliable estimates of the van der Waals coefficient $C_{6}$
directly by using expressions which relate this coefficient to the
frequency dependent dipole polarizabilities at imaginary
frequencies \cite{gisbergen1} which can be computed from TDDFT
using common functionals. We follow the latter route for the
calculation of these coefficients.

The paper is organized as follows: In section II, we discuss the
theoretical method and the expressions employed to calculate the
van der Waals coefficient $C_{6}$ from the frequency dependent
dipole polarizability. Results of our calculations are presented
in Section III.

\section{Method of Calculation}

In order to calculate the van der Waals coefficient $C_{6}$, we
make use of the Casimir-Polder expression which relates $C_{6}$ to
the frequency dependent dipole polarizability evaluated at
imaginary frequency. In accordance with this expression the
orientation averaged dispersion coefficient between two moieties
$A$ and $B$ is given by \cite{casimirpolder,stone}
\begin{equation}
C_{6}(A,B) = \frac{3}{\pi}\int_{0}^{\infty}d\omega\,
\bar{\alpha}_{A}(i\omega )\bar{\alpha}_{B}(i\omega )
\label{casimirpolder}
\end{equation}
where $\bar{\alpha}_{j}(i\omega)$ is the isotropic average dipole
polarizability of the $j$-th moiety and is given by
\begin{equation}
\bar{\alpha}_{j}(\omega) = \frac{\alpha^{j}_{xx}(\omega) +
\alpha^{j}_{yy}(\omega) + \alpha^{j}_{zz}(\omega)}{3}.
\end{equation}
In the above expression $\alpha_{xx}(\omega)$,
$\alpha_{yy}(\omega)$ and $\alpha_{zz}(\omega)$ are diagonal
elements of the dipole polarizability tensor. Therefore, the
calculation of dispersion coefficient $C_{6}$ involves determining
frequency dependent dipole polarizability tensor at a range of
imaginary frequencies followed by the evaluation of Eq.\ (1) by
numerical quadrature. For the determination of the frequency
dependent polarizability we use linear response theory based on
TDDFT, as already mentioned. For this work, the frequency
dependent polarizabilities of the dimers and  clusters were
obtained with the Amsterdam Density Functional (ADF) program
package \cite{adf}. We refer the reader to Ref.
\cite{gisbergen1,gisbergen2} for detailed description of the
method adopted in this package for obtaining frequency dependent
polarizabilities using ADF's \textsc{Response} module. This module
is restricted to the calculation of response properties of
closed--shell systems. Therefore, for the calculation of the
dynamic polarizability of the alkali-metal atoms Li, Na, and K, we
have employed an analytical TDDFT based sum-over-states (SOS)
expression for the polarizability \cite{fadda,jochen}. The SOS
approach for the frequency dependent polarizability requires the
computation of the excitation energies for the allowed transitions
and their corresponding oscillator strengths. These quantities can
also be obtained from TDDFT calculations \cite{casida}. For this
purpose we made use of the \textsc{Excitations} module of the ADF
program which allows the treatment of open-shell configurations
also. The SOS expression was evaluated by considering 85, 407, and
667 dipole--allowed excitations for the Li, Na, and K atom,
respectively. These are \emph{all} of the dipole allowed
excitations possible within the chosen STO basis sets (see below).
The SOS TDDFT results reported here are therefore equivalent to
the corresponding full linear response data for $\alpha (i\omega)$
(because of the large number of excitations using the SOS is
impractical for larger systems but quite feasible for atoms or
diatomics).

For the calculations of response properties by  TDDFT  one needs
to choose approximations for the XC potential and for the XC
response kernel.  The static XC potential is needed to calculate
the ground-state KS orbitals and their energies. The XC response
kernel $f_{XC}({\bf r},{\bf r'},\omega)$ determines the XC
contribution to the screening of an applied electric field. For
the XC kernel, we have used the adiabatic local density
approximation (ALDA) which was shown to be reasonably accurate for
atoms \cite{petersilka}. On the other hand, for the static XC
potential needed to calculate the ground-state orbitals and
energies, two different choices have been made. These are (i) the
standard local density approximation (LDA) as parameterized by
Vosko, Wilk and Nusair \cite{vwn} and (ii) a model potential,
called statistical average of orbital potentials (SAOP) which has
desirable properties both in the asymptotic and the inner regions
of a molecule \cite{gritsenko,schipper}. The results obtained by
these two XC potentials are compared in order to investigate the
effect of the XC potential on the dispersion coefficient C$_{6}$.
The SAOP yields improved results in particular for Rydberg
excitations where the asymptotic behavior of the XC potential
becomes important.

All calculations of the frequency dependent polarizabilities of
sodium clusters were carried out by using large Slater type
orbital (STO) basis sets. It is well known that for accurate
calculations of response properties it is necessary to have large
basis sets with both polarization and diffuse functions. For
alkali-metal atoms (Li, Na, and K) we choose the
quadruple--$\zeta$ triply polarized all electron even tempered
basis set ET-QZ3P-3DIFFUSE from the ADF basis set library which
has three sets of diffuse functions. On the other hand, for dimers
of alkali-metal atoms and clusters of sodium atoms a slightly
smaller yet accurate all electron basis set ET-QZ3P-2DIFFUSE has
been used to reduce the computational time and cost. The
application of highly flexible atomic orbital basis sets with
diffuse functions often leads to the problem of linear
dependencies. Such problem have been circumvented by removing
linear combinations of functions corresponding to small
eigenvalues of the overlap matrix. We expect that the size of the
chosen basis set will make our results quite close to the
basis-set limit.

The Casimir-Polder integral Eq. (\ref{casimirpolder}) has been
evaluated by employing a thirty point Gauss-Chebyshev quadrature
scheme as described in Ref. \cite{rijks}. The convergence of the
results have been checked by comparing the results for increasing
numbers of frequency points.

In order to perform the TDDFT calculations of the frequency
dependent polarizabilities of  the dimers and clusters, we needed
to choose their ground-state geometries. For the dimers, we used
experimental bond lengths 2.6725 \AA\ for Li$_{2}$, 3.0786 \AA\
for Na$_{2}$, and 3.923 \AA\ for K$_{2}$ as in Ref.
\cite{pacheco1}. On the other hand, for larger clusters (4- to
20-atom clusters), we use structures which have been obtained via
geometry optimizations employing a triple-$\zeta$ STO basis with
two added polarization functions (TZ2P basis from the ADF basis
set library) along with the Becke-Perdew (BP86) XC potential
\cite{becke2,perdew2}. This XC potential is known to yield
reliable geometries. All the optimizations have been carried out
with the convergence criteria for the norm of energy gradient and
energy, fixed at $10^{-4}$ atomic units (a.u.) and $10^{-6}$ a.u.,
respectively. In Ref.\cite{banerjee2}, we employed these
geometries to calculate C$_{6}$ for sodium clusters. In case of a
cluster having more than one isomers, we choose the one possessing
the lowest energy for our calculations of the dipole
polarizability.

\section{Results and Discussion}

Before proceeding with the detailed discussion on our results for
the dispersion coefficients we shall first assess the accuracy of
the TDDFT based analytical SOS expression and of the SAOP XC
potential (compared to the LDS) in predicting  the dynamic
polarizabilities of alkali-metal atoms. To this end we have
calculated the frequency dependent linear polarizabilities $\alpha
(\omega)$ for Li, Na and K atoms over wide range of frequencies
and compared with theoretical results available in the literature.
In Figs. 1-3 we display the frequency dependent linear
polarizability of Li, Na and K atoms, respectively, for real
frequencies. In each figure we consider two different ranges of
frequencies. In part $(a)$ a frequency range spanning $\omega =
0-0.035$ Hartree (a.u.) has been chosen to compare our the results
with those of Refs. \cite{pipin} (Hylleraas approach),
\cite{kobyashi} (Moller-Plesset perturbation theory), and
\cite{hibbert} (CI approach). A wider frequency range is chosen in
part $(b)$ of each figure, which is similar to the one considered
in Ref. \cite{safronova} for the calculation of dynamic
polarizabilities of alkali-metal atoms by using a combination of
the random-phase approximations for the core electrons and the
contribution of valence electrons were obtained using very
accurate oscillator strength and transition energy data. For
clarity we display the results for each atom in two separate
graphs as the frequency mesh used for the above two ranges are
quite different. It can be clearly seen from these figures that
LDA results for the frequency dependent polarizabilities for all
the atoms are always lower than the SAOP data throughout the whole
frequency range. Both the static values of the polarizabilities
and their frequency dependence are underestimated by LDA XC
potential. This is consistent with the fact that the LDA potential
fails to exhibit the correct behavior both in the inner and
asymptotic regions of the molecule -  which is required for
accurate determination of the frequency dependent dipole
polarizability. In comparison, the SAOP XC potential possesses
much improved properties both in the asymptotic and the inner
region of a molecule and consequently it is expected that the
results obtained with this potential will be in better agreement
with the other accurate theoretical results available in the
literature. The improvement of SAOP results over the LDA results
obtained with the same STO basis are clearly elucidated in Figs.
1-3. We also observe from Figs. 1 -3 that the frequency dependent
polarizability obtained with SAOP XC potential are still slightly
lower than the results of Refs.
\cite{pipin,kobyashi,hibbert,safronova} except for the case of K
atom. The differences between SAOP and the other results shown in
Figs. 1-3 are uniform over whole frequency range for Li and Na
atoms. A SAOP XC kernel has not yet been implemented, therefore we
are unable to make a direct comparison but it is likely that the
ALDA approximation for the XC kernel, along with differences in
the basis sets applied, is responsible for the remaining small
differences between our and the literature data.

Having demonstrated the applicability and assessed the accuracy of
the analytical SOS expression within TDDFT for the calculation of
frequency dependent polarizabilities of alkali atoms Li, Na, and
K, we now proceed with the discussions of the results for $C_{6}$.
First we present the results for $C_{6}$ between different pairs
of alkali-metal atoms obtained by employing the LDA and SAOP XC
potentials. These results are presented in Table I and compared
with other theoretical results available in the literature
\cite{marinescu,muller,speisberg,maeder,stanton,rerat}. In Refs.
\cite{muller,speisberg,maeder} the calculations were performed by
employing a configuration interaction (CI) method for the valence
orbitals. The core electrons were treated using a pseudopotential
approach. Other quantum chemical methods such as a couple cluster
approach \cite{stanton} and an \textit{ab initio} time dependent
gauge invariant method coupled with a CI method \cite{rerat} have
been employed to calculate the dynamic polarizability at imaginary
frequencies. On the other hand, by constructing precise
single-electron  model potentials to represent the motion of the
valence electron in the field of the closed--shell alkali-metal
positive-ion core the calculation of frequency dependent
polarizabilities at imaginary frequencies and C$_{6}$ of
alkali-metal atoms have been performed in Ref. \cite{marinescu}.

First we note that, as it was the case for the polarizability, the
C$_{6}$ coefficients obtained with SAOP are systematically higher
than the corresponding LDA data. The comparison of other
theoretical results compiled in Table I with the corresponding
SAOP data clearly shows that SAOP value of C$_{6}$ for Li-Li
interaction is slightly higher (around $2.5\%$) relative to  all
the results presented in Table I except for the data of Ref.
\cite{stanton}. As a matter of fact for all the diatom pairs
results of Ref \cite{stanton} are higher than all other results
displayed in Table I.  In contrast to the Li-Li case for Na-Na and
K-K interactions SAOP results are slightly lower than the
corresponding numbers obtained with other theoretical methods
except for the results of Ref. \cite{marinescu} and Ref.
\cite{muller} for Na-Na and K-K cases respectively. In particular
for Na-Na interaction the SAOP value of C$_{6}$ differs slightly (
higher by around 1 atomic unit) as compared to the data of Ref.
\cite{marinescu}. For heteronuclear cases of Li-Na and Li-K
interactions the agreement between SAOP and the other theoretical
results presented in Table I is quite good. For Na-K interaction,
SAOP number for C$_{6}$ is lower than all the results and a
maximum difference (around $6\%$) is found with result of Ref.
\cite{maeder}. These results then clearly demonstrate that the
TDDFT approach used here (with the SAOP potential) is capable of
predicting quite accurate C$_{6}$ of alkali-metal diatoms as these
results lie well within the range of values produced by other
correlated wavefunction based methods. Mostly, our results compare
very well with other data listed in Table I.

We now present C$_{6}$ for the interactions between alkali-metal
atoms and alkali dimers. These results are displayed in Table II
and compared with the results of Ref. \cite{speisberg}. We mention
here that Ref. \cite{speisberg} employed slightly different values
of bond lengths for the dimers of alkali atoms in comparison to
the ones used in our calculations. Like for our other results, we
find that the LDA values for the atom-dimer interactions are again
systematically lower than the corresponding SAOP results. For both
the homonuclear and heteronuclear cases the SAOP results are a
little higher than the corresponding data of Ref.
\cite{speisberg}. The largest difference of around $8\%$ is found
for Li-Li$_{2}$ interaction whereas the difference is the smallest
for Na-Na$_{2}$ (around $1\%$). In general it appears that the
differences between our SAOP results and the data of Ref.
\cite{speisberg} for the interaction between alkali atoms and the
Li$_{2}$ dimer are somewhat larger than the corresponding
differences for Na$_{2}$ and K$_{2}$. For example, the difference
between the results for K-Na$_{2}$ and K-K$_{2}$ are of the order
of 3$\%$ whereas for K-Li$_{2}$ it is around 7$\%$. Similarly for
Na-Na$_{2}$ and Na-K$_{2}$ the discrepancy between the results are
just 1$\%$ while it is around 5$\%$ for Na-Li$_{2}$ case. Overall,
however, the agreement between our data and those of Ref.\
\cite{speisberg} is very encouraging.

Finally, we discuss the results for $C_{6}$  pertaining to the
Li-Na$_{n}$, Na-Na$_{n}$, and K-Na$_{n}$ interactions.  As
mentioned before an experiment involving the measurement of the
integral scattering cross section in low energy collisions between
neutral sodium clusters Na$_{n}$ ( $2\leq n \leq 20$) and the
alkali atoms Li. Na and K  was performed by Kresin et al.
\cite{kresin1}. It has been shown in Ref. \cite{kresin1} that the
values of $C_{6}$ calculated from the London dispersion formula
given by
\begin{equation}
C_{6} =
\frac{3}{2}\alpha_{A}(0)\alpha_{B}(0)\frac{\omega_{A}\omega_{B}}{\omega_{A}
+ \omega_{B}},
\label{london}
\end{equation}
with $\omega_{i}$ and $\alpha_{i}(0)$ denoting the characteristic
frequency and static polarizability of the collision partners
yield results for the integral scattering cross sections which
show a good agreement with the experimental data. For details on
the values of dipole transition frequencies and static
polarizabilities employed to calculate $C_{6}$ we refer the reader
to Ref. \cite{kresin1}. In this paper, we compare the
London--formula based $C_{6}$ coefficients with our
first--principles results (using the SAOP data). In Figs. 4a, 4b
and 4c we display $C_{6}$ coefficient for the pairs $Li-Na_{n}$,
$Na-Na_{n}$, and $K-Na_{n}$, respectively, as functions of the
number of atoms present in the cluster. It can be clearly seen
from Fig. 4 that for all the three cases of atom-cluster
interactions, the TDDFT and London formula results for C$_{6}$ are
quite close to each other for magic--number clusters containing 2,
8, and 20 sodium atoms. For other pairs, the  match between the
two data are still reasonably good. The largest differences are
found for the pairs $Li-Na_{16}$, $Na-Na_{16}$, and $K-Na_{16}$.
These results are consistent with the fact that the London's
formula does not take anisotropic nature of the clusters into
account. As a result of this  the magic-number clusters which show
less anisotropy than the non-magic ones \cite{banerjee2} are well
described by London's formula. The overall agreement of the
$C_6$-coefficients obtained here from the TDDFT computations with
those derived from the London formula may be attributed to the
fact that for alkali-metal atoms and sodium clusters
\cite{kresin2,deheer,madjet,kummel} the optical absorption spectra
exhibit one strong resonance carrying essentially all the
transition strength, which is also the basic assumption made in
deriving the London formula. This is confirmed by our
first--principles computations, i.e.\ our results show that the
approximate London dispersion formula is indeed well suited for
calculating the dispersion coefficient C$_{6}$ for interactions
between alkali atoms and magic number sodium clusters.

\section{Summary and Conclusions}

This paper is devoted to the calculation of long-range van der
Waals coefficient $C_{6}$ for the interactions between
alkali-metal atoms Li, Na, and K and their dimers, and sodium atom
clusters containing an even number of atoms ranging from 2 to 20.
The calculations were performed by employing all-electron TDDFT
methods. The van der Waals coefficient has been obtained by using
the Casimir-Polder expression which needs frequency dependent
dipole polarizabilities of the two interacting species as input.
The frequency dependent polarizability of the atoms were obtained
by employing a TDDFT--based analytical complete SOS expression
while for all other systems (dimers and clusters) direct linear
response theory within TDDFT has been used. The calculations were
performed by using a model XC potential (SAOP) having the correct
behavior in the asymptotic region (as well as improved behavior in
the valence and core regions of the molecules, compared to LDA).
The calculations were carried out with one of the largest STO
basis sets available in ADF basis set library; therefore, the
results are expected to be reasonably close to the complete basis
limit. In this paper the performance of the SAOP and LDA XC
potentials for the calculations of the frequency dependent
polarizability of alkali-metal atoms have been compared against
other theoretical results available in the literature. We found
that SAOP results are in much better agreement with published data
than the LDA results and compare well with results obtained by
employing \textit{ab initio} correlated wave-function based
methods. Motivated by these encouraging results we then carried
out calculations of the coefficient C$_{6}$ for different
atom/diatomic - cluster pairs as mentioned above and compared our
results with other theoretical data where available. The results
presented in this paper clearly showed that TDDF with the the SAOP
XC potential performs very well in the computation of van der
Waals coefficient C$_{6}$ also. For atom-atom and atom-dimer
interactions, we found that SAOP results are quite close to the
data available in literature obtained by employing various
correlated wave-function based methods. As no theoretical results
are available for C$_{6}$ between alkali-metal atoms and sodium
clusters, we made comparisons of our TDDFT based results with
those obtained by Kresin et al.\cite{kresin1} by using the London
dispersion formula which is valid under the approximation that the
absorption spectra exhibits single strong resonance peak at an
effective frequency. These comparisons clearly reveal that the
overall agreement is quite good and specially for magic--number
clusters with 2, 8, and 20 atoms the approximate London formula
yields values for C$_{6}$ which are very close to our
first--principles results. We attribute the agreement of the
results obtained with TDDFT and with the London formula to the the
fact that a single strong resonance peak dominates the absorption
spectra of the alkali-metal atoms and sodium clusters.

\acknowledgments{ A. B. wishes to thank Mr. Pranabesh Thander of
RRCAT Computer Centre for his help and support in providing us
uninterrupted computational resources and also for smooth running
of the codes and Dr. Aparna Chakrabarti for her help with the
geometry optimization. J.\ A.\ acknowledges support from the
Center of Computational Research at SUNY Buffalo and is grateful
for financial support from the CAREER program of the National
Science Foundation. It is a pleasure to thank Prof. Vitaly Kresin
for valuable suggestions and making his numbers for $C_{6}$
available to us.}

\clearpage

\newpage

\section*{Figure captions}

\noindent {\bf Fig.1} Comparison of linear polarizability  $\alpha
(\omega)$ ( in Hartree atomic units) of the Li atom as a function
of frequency obtained by different methods: TDDFT (SAOP and LDA,
this work), Moller-Plesset perturbation theory (Ref.
\cite{kobyashi}), Hyleraas wavefunction approach (Ref.
\cite{pipin}), and Random-phase approximation and SOS expression
(Ref. \cite{safronova}). The lines joining the points were added
to guide the eye.

\noindent {\bf Fig.2} Same as Fig. 1 but for Na atom. TDDFT (SAOP
and LDA, this work),  Moller-Plesset perturbation theory (Ref.
\cite{kobyashi}), CI approach (Ref. \cite{hibbert}), and
Random-phase approximation and SOS expression (Ref.
\cite{safronova}). The lines joining the points were added to
guide the eye.

\noindent {\bf Fig.3} Same as Fig. 1 but for K atom. TDDFT (SAOP
and LDA, this work), Moller-Plesset perturbation theory (Ref.
\cite{kobyashi}), and Random-phase approximation and SOS
expression (Ref. \cite{safronova}). The lines joining the points
were added to guide the eye.

\noindent {\bf Fig.4} Comparison of TDDFT  and London formula
based  results for the van der Waals coefficient $C_{6}$ ($\times
10^{-3}$) corresponding to the alkali-atom-cluster pairs (a)
$Li-Na_{n}$, (b) $Na-Na_{n}$, and (c) $K-Na_{n}$. The numbers for
$C_{6}$ are in atomic units. The London formula results
(represented by solid circles) were taken from Ref.
\cite{kresin1}. The lines joining the points were added to guide
the eye.

\clearpage
\newpage
\begin{table}
\caption{Results for C $_{6}(\times 10^{-3}$) between different
pairs of alkali-metal atoms, in atomic units} \tabcolsep=0.1in
\begin{center}

\begin{tabular}{|c|c|c|c|c|c|c|}\hline
Source  & Li-Li & Na-Na & K-K & Li-Na & Li-K & Na-K\\
\hline
Present (LDA) & 1117.65 & 1243.51 & 3320.71 & 1176.73 & 1922 & 2019  \\
Present (SAOP) & 1426 & 1473 & 3590 & 1448 & 2257 & 2288  \\
Ref. \cite{speisberg} & 1385 & 1527 & 3637 & 1452 & 2238 & 2336  \\
Ref. \cite{muller} & 1386 & 1518 & 3574 & 1448 & 2219 & 2309  \\
Ref. \cite{maeder} & 1389 & 1540 & 3945 & 1460 & 2333 & 2443  \\
Ref. \cite{marinescu} & 1388 & 1472 & 3813 & 1427 & 2293 & 2348  \\
Ref. \cite{stanton}  & 1439 & 1639 & 4158 & 1532 & 2441 & 2595 \\
Ref. \cite{rerat}  & 1419 & 1554 & - & 1479 & - & - \\
\hline
\end{tabular}
\end{center}
\end{table}

\begin{table}
\caption{ Results for C $_{6}(\times 10^{-3}$) between different
alkali-metal atoms and alkali dimers, in atomic units. For each
pair the first and second row numbers are obtained with LDA and
SAOP XC potentials respectively. The number in the parenthesis for
each pair is taken from Refs. \cite{speisberg}} \tabcolsep=0.1in
\begin{center}
\begin{tabular}{c|ccc}
$\frac{DIMERS}{ATOMS}$  & Li$_{2}$ &  Na$_{2}$ &  K$_{2}$ \\
\hline
Li & 1696 & 2018 & 3338   \\
   & 2108 & 2513 & 3967   \\
   &(1935)& (2394) & (3791)  \\
Na & 1794 & 2138 & 3516 \\
   & 2148 & 2562 & 4029 \\
   & (2039)  & (2524) &(3966) \\
K  & 2910   & 3459   & 5759  \\
    & 3327   & 3969    & 6302  \\
    & (3102)    & (3838)   & (6144 )   \\
\end{tabular}
\end{center}
\end{table}

\newpage
\clearpage
\begin{figure}
\begin{center}
\includegraphics[height=10cm,width=15cm]{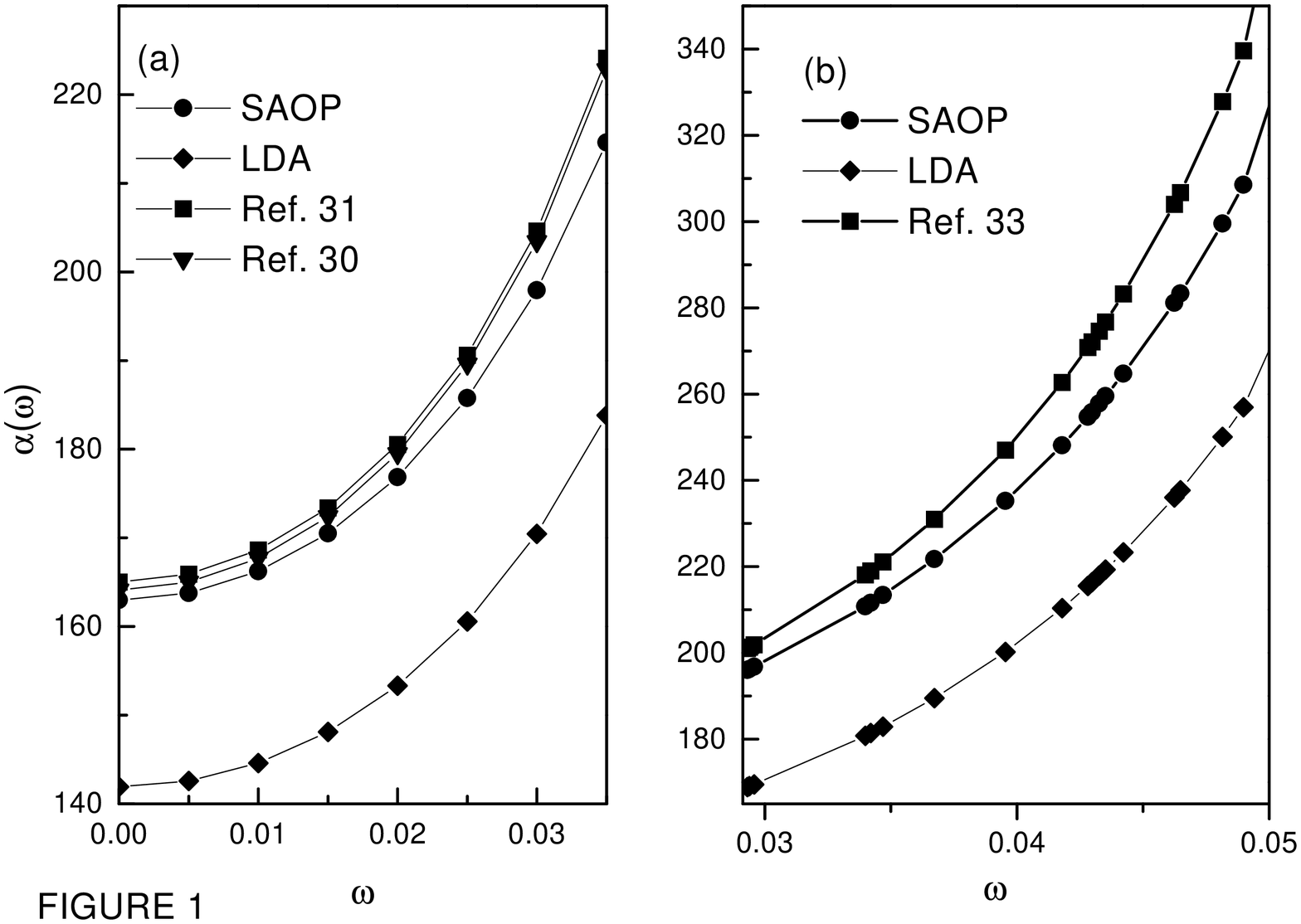}
\label{fig1}
\end{center}
\end{figure}
\begin{figure}
\begin{center}
\includegraphics[height=10cm,width=15cm]{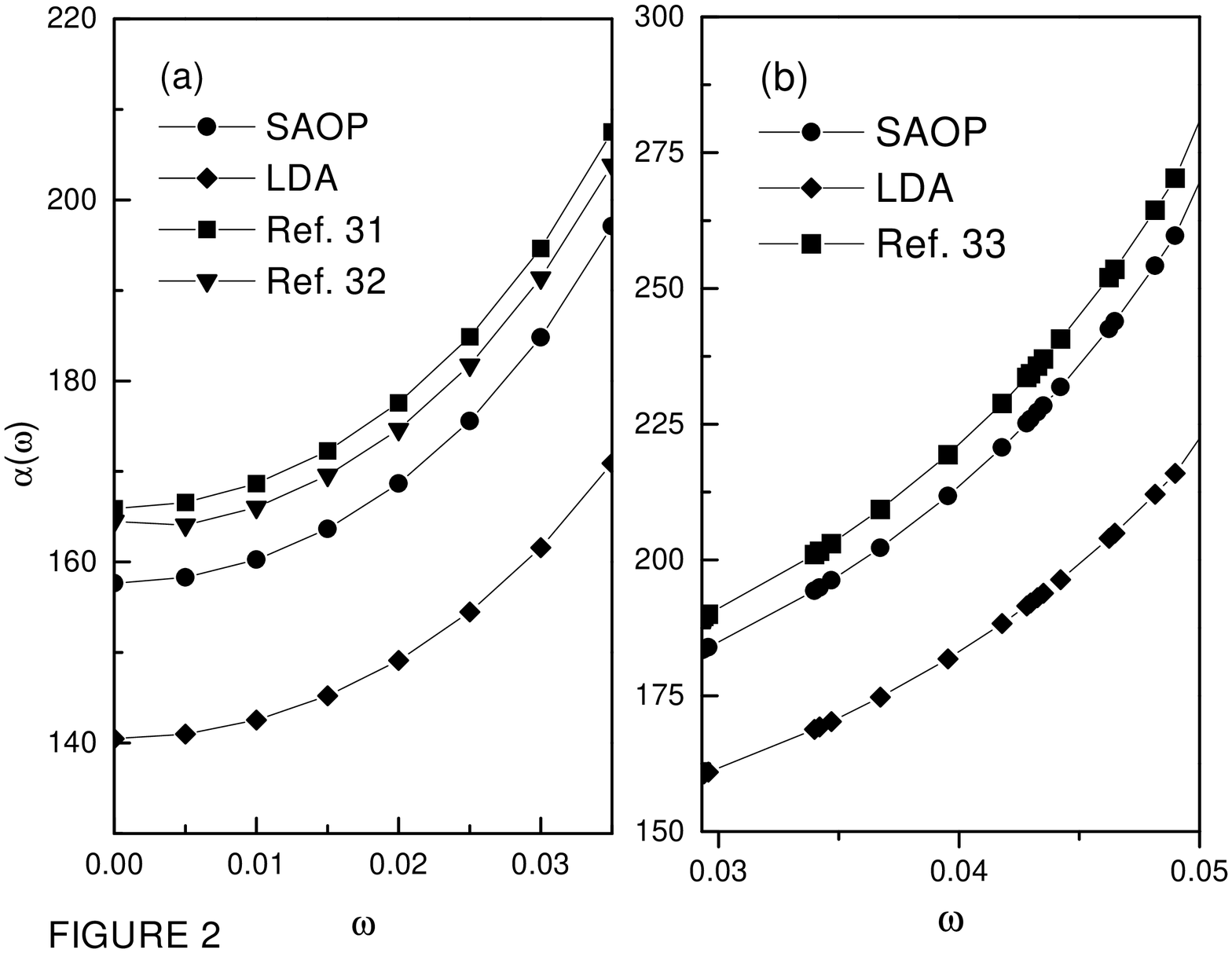}
\label{fig1}
\end{center}
\end{figure}
\begin{figure}
\begin{center}
\includegraphics[height=10cm,width=15cm]{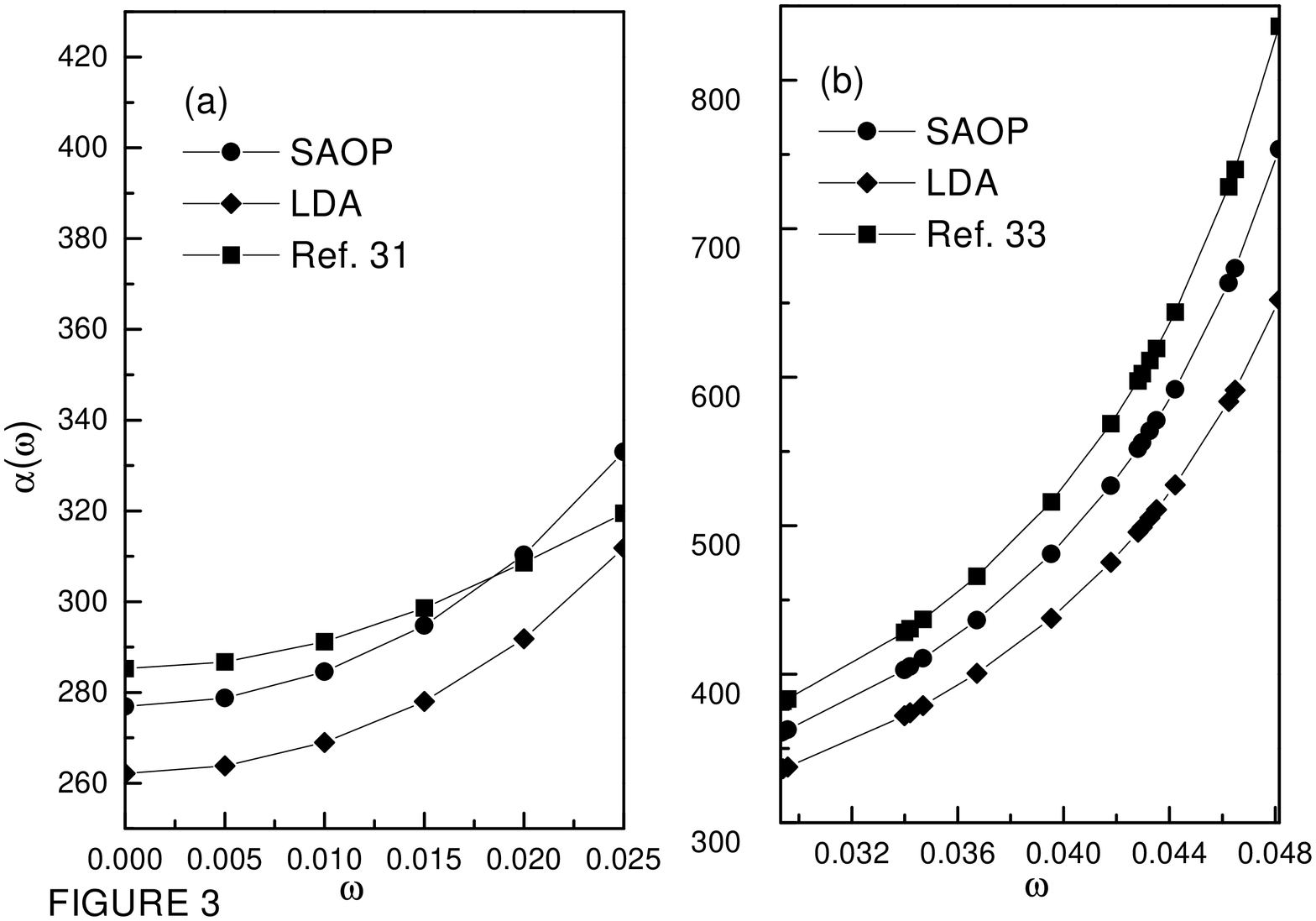}
\label{fig1}
\end{center}
\end{figure}
\begin{figure}
\begin{center}
\includegraphics[height=25cm,width=15cm]{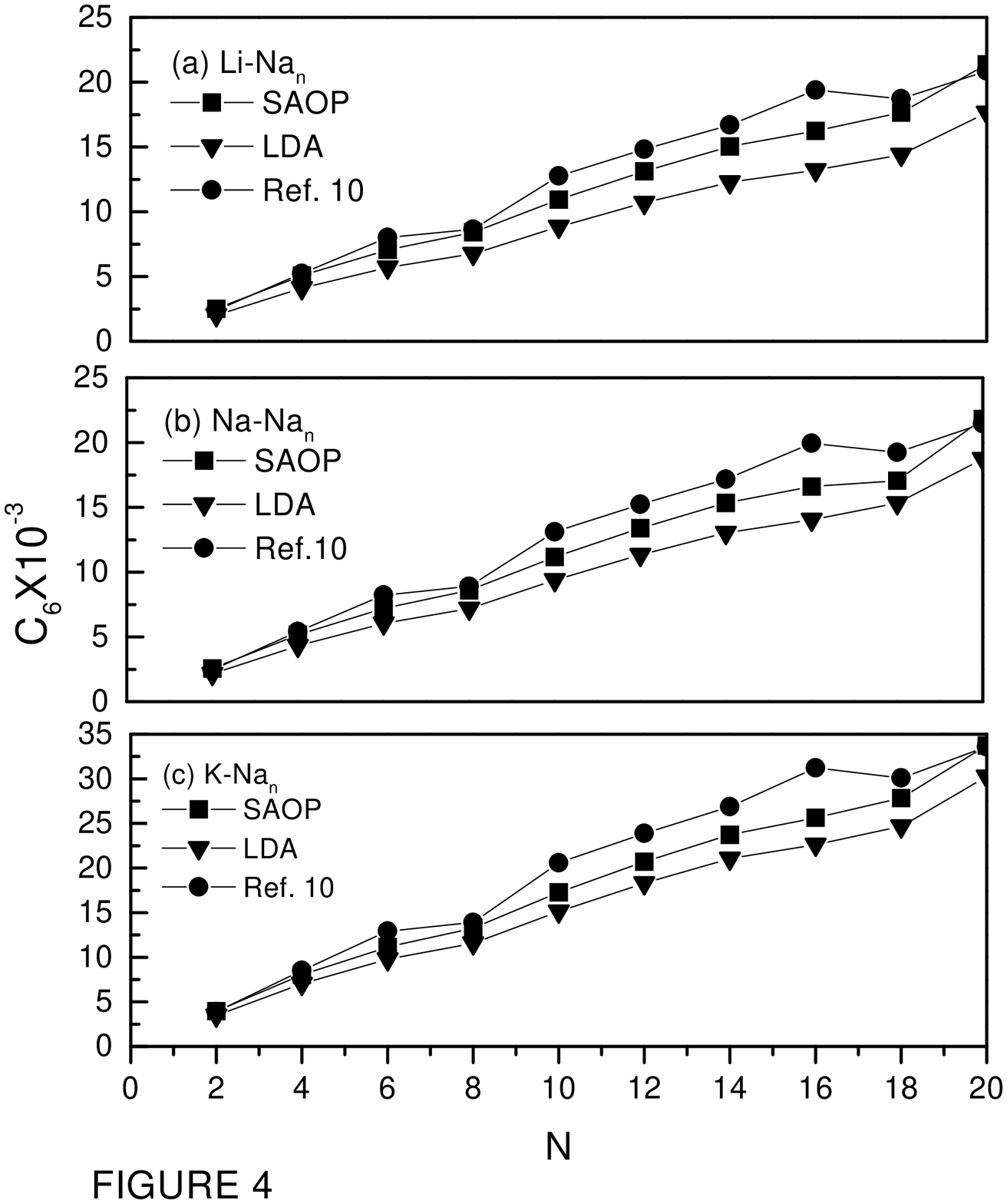}
\label{fig1}
\end{center}
\end{figure}
\end{document}